\documentstyle[psfig,sprocl]{article}

\def\ga{\mathrel{\mathchoice {\vcenter{\offinterlineskip\halign{\hfil
$\displaystyle##$\hfil\cr>\cr\sim\cr}}}
{\vcenter{\offinterlineskip\halign{\hfil$\textstyle##$\hfil\cr>\cr\sim\cr}}}
{\vcenter{\offinterlineskip\halign{\hfil$\scriptstyle##$\hfil\cr>\cr\sim\cr}}}
{\vcenter{\offinterlineskip\halign{\hfil$\scriptscriptstyle##$\hfil\cr>\cr\sim\cr}}}}}

\title{Evolution of Galaxy Clustering}

\author{J.S.Bagla}

\address{Institute of Astronomy, University of Cambridge, Madingley Road, \\
Cambridge CB3 0HA, U.K. \ \  E-mail: jasjeet@ast.cam.ac.uk}

\begin{document}

\maketitle\abstracts{We show that the galaxy correlation
function does not evolve in proportion with the correlation function
of the underlying mass distribution.  Earliest galaxies cluster very
strongly and the amplitude of the galaxy correlation function {\it
decreases} from this 
large value.  This continues till the average peaks have collapsed,
after which, the galaxy correlation function does not evolve very
strongly.}

\section{The Halo Grail}

We begin by addressing the simpler problem of halo correlation.  In
later sections we will test the model presented here using N-Body
simulations.  Lastly, we will comment on applying the results for halo
clustering to galaxy clustering and discuss some of its implications.

Consider the distribution of halos of mass M, before typical halos of
this mass have collapsed.  To quantify this, we first 
define a {\it bias} parameter $\nu(M) = \delta_c/\sigma(M)$, where
$\sigma(M)$ is the (linearly extrapolated) {\it rms} dispersion in
density at mass scale $M$ and $\delta_c=1.68$ is the linearly
extrapolated density contrast at which halos are expected to
virialise.  We can write the linear correlation function of these
halos, for $\nu \gg 1$ and $\frac{\xi(M,r)}{\xi(M,0)} \ll 1$, as \cite{bbks}
\begin{equation}
\xi_H(M,r) = \exp\left[ \nu^2 \frac{\xi(M,r)}{\xi(M,0)} \right] - 1
\label{halocorr} 
\end{equation}
Here $\xi(M,r)$ is the correlation function of the density field
smoothed at mass scale $M$, evaluated at scale $r$ and $\xi_H$ is the
correlation function of halos.  It follows from this expression
that the halo correlation function is much larger than the mass
correlation function in the range of scales where $\nu^2
\frac{\xi(M,r)}{\xi(M,0)} \ga 1$.  The evolution of $\xi_H$ for these
scales is controlled by the decreasing function $\nu$.  {\it
Therefore, at early times, the amplitude of correlation 
function of halos is a decreasing function of time.}
Eqn.(\ref{halocorr}) gives only the linearly extrapolated correlation
function for halos.  However, the qualitative behaviour, being
exponentially strong, should survive the non-linear evolution.   

At later epochs, when $\nu(M) \approx 1$, eqn.(\ref{halocorr}) is no
longer valid. By this time, most halos of mass $M$ have collapsed.
In hierarchical models, these halos merge and give rise to more
massive halos.  As gravity brings halos closer, the
halo correlation function increases. However, the rate of growth of
correlation function will be slow as {\it anti-biased} halos
continue to collapse for some time. 

\section{Simulations}

The ideas outlined above are applicable to all models, we tested
these using simulations of the SCDM model.  We used a $128^3$ PM
simulation with box size of $90$h$^{-1}$Mpc.  We normalised the power
spectrum to reproduce the present day cluster abundance ($\sigma_8 =
0.6$).  Halos were identified with the friends of friends (FOF)
algorithm, with a threshold linking length of $0.2$ grid lengths. 

In principle, we should use halos with mass in a narrow range to study
the evolution of halo correlation function.  However, two reasons
force us to use a different strategy.  
(1) The FOF algorithm is known to link together dynamically distinct
halos.  This leads to an incorrect estimate of the halo mass and the
number of close pairs of halos.  
(2) Finite mass resolution in numerical simulations leads to the
over merging problem \cite{overmerging} and this makes it difficult to
estimate the number of halos that survive inside bigger halos.
Both of these lead to an underestimate of the halo correlation
function.  Therefore, we do not consider halos as being one unit each.
We assign a weight, proportional to the mass, to each halo. 
Operationally, this is equivalent to computing the correlation
function of particles contained in these halos.  This strategy is also
appropriate for studying galaxy correlation function
as galaxies are known to survive inside clusters of galaxies.
Although we ignore the large range in masses of galaxies by
using this particular method, we include a realistic contribution
of very massive halos.

We describe the clustering properties using the averaged two
point correlation function $\bar\xi$, defined as 
\begin{equation} 
\bar\xi(r) = \frac{3}{r^3} \int\limits^r x^2 \xi(x) dx 
\end{equation}
where $\xi$ is the two point correlation function.   

To quantify the differences in evolution of halo distribution and mass
distribution, we have plotted the averaged correlation function for
these in fig.1.  The left panel shows the growth of
the mass correlation function and the right panel shows the
evolution of the halo correlation function (Minimum halo mass for this
simulation is $2\times 10^{12} M_\odot$).  This figure shows that the
clustering in mass increases monotonically.  In contrast, the
amplitude of the halo correlation function is high at early times and
decreases rapidly up to $z=1$.  It starts increasing again after $z
\simeq 0.5$ which corresponds closely to $\nu(M) \simeq 1$.  The halo
correlation function varies very little between $z=1$ and $z=0$.  The
value of $\nu(M)$ that corresponds to the minimum amplitude of halo
correlation function depends on the local index of the power spectrum
\cite{galclus} and is smaller for negative indices. 

\begin{figure}
\psfig{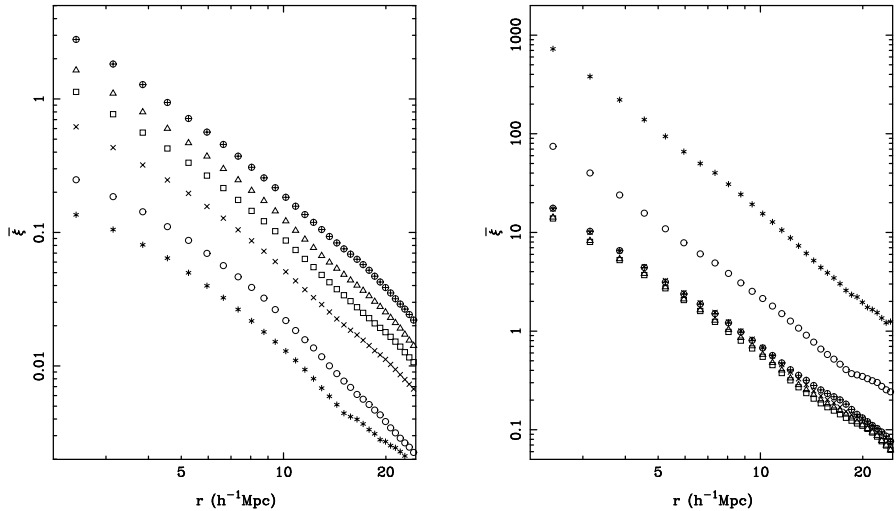}
\caption{The left panel shows the evolution of clustering in
the total mass distribution, $\bar\xi$ increases monotonically.
Symbols are $\ast$ ($z=3$), $\bigcirc$  ($z=2$), $\times$ ($z=1$),
$\Box$  ($z=0.5$), $\triangle$ ($z=0.25$) and $\oplus$ ($z=0$).  The
halo correlation function, shown in right panel, decreases from its
initial high value at $z=3$, slows down after $z=1$ and starts
increasing after $z=0.5$.} 
\end{figure}

\section{Discussion}

In order to apply the results given above to clustering of galaxies,
we must incorporate effects of the mass function of galaxies.  The
amplitude of halo correlation function starts increasing earlier for
low mass halos.  As there are many more low mass halos than high mass
ones, the epoch at which the amplitude of galaxy correlation function
starts increasing will depend on the mass of the smallest galaxies
that can be seen at high redshifts \footnote{Identifying the {\it
same}\/ population of galaxies at different redshifts is also a
problem.}.  Simulations that combine semi-analytic models of galaxy 
formation with gravitational clustering can be used to compute galaxy
correlation function and for the models where this has been done, the
galaxy correlation function follows the same pattern as the halo
correlation function.\cite{tcdm}  Irrespective of the detailed
evolution, we can conclude, that at high redshifts, galaxies cluster
much more strongly than the underlying mass distribution.  Thus, the
observed clustering of galaxies at high redshifts \cite{wallobs} is
not a strong constraint for most models of structure formation.\cite{walls} 

Some other implications of strong clustering at high redshifts 
are: (1)~The evolution of galaxy clustering is not a good indicator of
cosmological parameters.  (2)~The shape of the galaxy correlation
function is different from the shape of the mass correlation function.
Therefore galaxy correlation function is not a very good indicator of
the initial power spectrum.  (3)~Sources responsible for reheating
and reionisation of the IGM will have a very non-uniform
distribution.  This will lead to a patchy structure at early epochs.
The scale of patchiness, which may be observable,\cite{tomo} can be
used to constrain galaxy formation scenarios.  (4)~Formation of many
ionising sources in a small region will increase the temperature of
the IGM and inhibit collapse of low mass halos in these
regions.\cite{supress}  Therefore, the mass function of galaxies near
and away from these ionising centres will be different.  (5)~If
quasars form preferentially in high mass halos then they should show
stronger clustering than galaxies.  (Recent estimates of quasar
correlation function show that it is stronger than the galaxy
correlation function.\cite{qsocorr1}$^{\!,\,}$\cite{qsocorr2})  A
comparison of the two, and their evolution, can be a useful indicator
of the prevalence of AGN activity in galaxies. 

\section*{Acknowledgements}

I acknowledge the support of PPARC fellowship at the Institute of
Astronomy.  I thank the organisers of the {\it Large Scale Structure:
Tracks and Traces}\/ for their kind hospitality.

\label{lastpage}

\end{document}